\documentclass[conference]{IEEEtran}
\usepackage{graphicx}
\usepackage{caption}

\begin{document}
\title{Analysis of boundary point (break point) in Linear Delay Model for nanoscale VLSI standard cell library characterization at PVT corners} 
\author{Gaurav Kumar Agarwal \\
	Department of Electronics and Communication Engineering, Indian Institute of Technology Roorkee \\
	Roorkee, India (gauravagru@gmail.com)}
\maketitle

\begin{abstract}
In VLSI chip design flow, Static Timing Analysis (STA) is used for fast and accurate analysis of data-path delay. This process is fast because delay is picked from Look Up Tables (LUT) rather than conventional SPICE simulations. But accuracy of this method depends upon the underlying delay model with which LUT was characterized. Non Linear Delay Model (NLDM) based LUTs are quite common in industries[\ref{nldm}]. These LUT requires huge amount to time during characterization because of huge number of SPICE simulations done at arbitrary points. To improve this people proposed various other delay models like alpha-power[\ref{alpha-model}] and piecewise[\ref{piecewise-model}] linear delay models. Bulusu et. al.[\ref{bulusu}] proposed Linear Delay Model(LDM) which reduces LUT generation time to 50 percent. LDM divides delay curve w.r.t input rise time($t_{rin}$) into two different region one is linear and other is non-linear. This boundary point between linear and non-linear region was called break point ($t_{rb}$). Linear region will be done if we simulate at only two points. This advantage will be possible by having knowledge of this break point at various PVT corners. In this paper, We will analyze this break point and will give a formula to find out this at various PVT corners. Knowledge about $(t_{rb})$ will restrict LUT simulations only in non-linear region and will help us in saving huge amount of time during LUT characterization.
\end{abstract}

\subsection*{Keywords}
Static Timing Analysis, Look Up Table characterization, data path delay estimation, linear delay model

\section{Introduction}
In Static Timing Analysis, hold and setup time violations have to be validated. For a combinational circuit, setup time of the flip-flop puts the constraint on upper limit of the delay while hold time puts a constraint on lower limit. This makes accurate estimation of data-path delay necessary.

There are various ways to estimate this data-path delay but LUT based delay estimation is fastest one. An LUT holds delay values of a circuit at various corners of input rise time($T_R$)\footnote{$t_{rin}$ is input rise time from 20\% to 80\% of transition and $T_R$ is with respect to 0 to 100\% of transition.} and output load ($C_L$). For different type of standard gate like NAND, NOR, NOT a different LUTs are characterized and this process is replicated at various Process, Supply Voltage and Temperature (PVT) corners. .

These $T_R$ and $C_L$ points are chosen arbitrarily or at uniform intervals. Each point requires an SPICE simulation. If multiplied for all values of  $T_R$ and $C_L$, it takes a huge amount of time. Bulusu et al.[\ref{bulusu}] used the fact that delay varies linearly up to some extent with $T_R$ and $C_L$ and could reduce the number of required simulations greatly. They exploited this linear variation by choosing simulation points of  $T_R$ and $C_L$  only in non-linear region. Thereby getting away with only two simulations point in linear region.

In this paper, We will summarize the linear delay model in Section II. In Section III, We will calculate $t_{rb}$ for various $C_L$, Supply Voltage ($V_{DD}$), Temperature and in Section IV will verify these calculations using HSPICE simulation. Finally in section V, We will give a model for $t_{rb}$ that will enable us to extract out linear region at various Voltage and Temperature corners. This in turn will reduce number of required simulations at any PVT corner. 

\section{Summary of Linear Delay Model[\ref{bulusu}]}

\begin{figure}[ht!]
	\centering
	\includegraphics[width=3.0in]{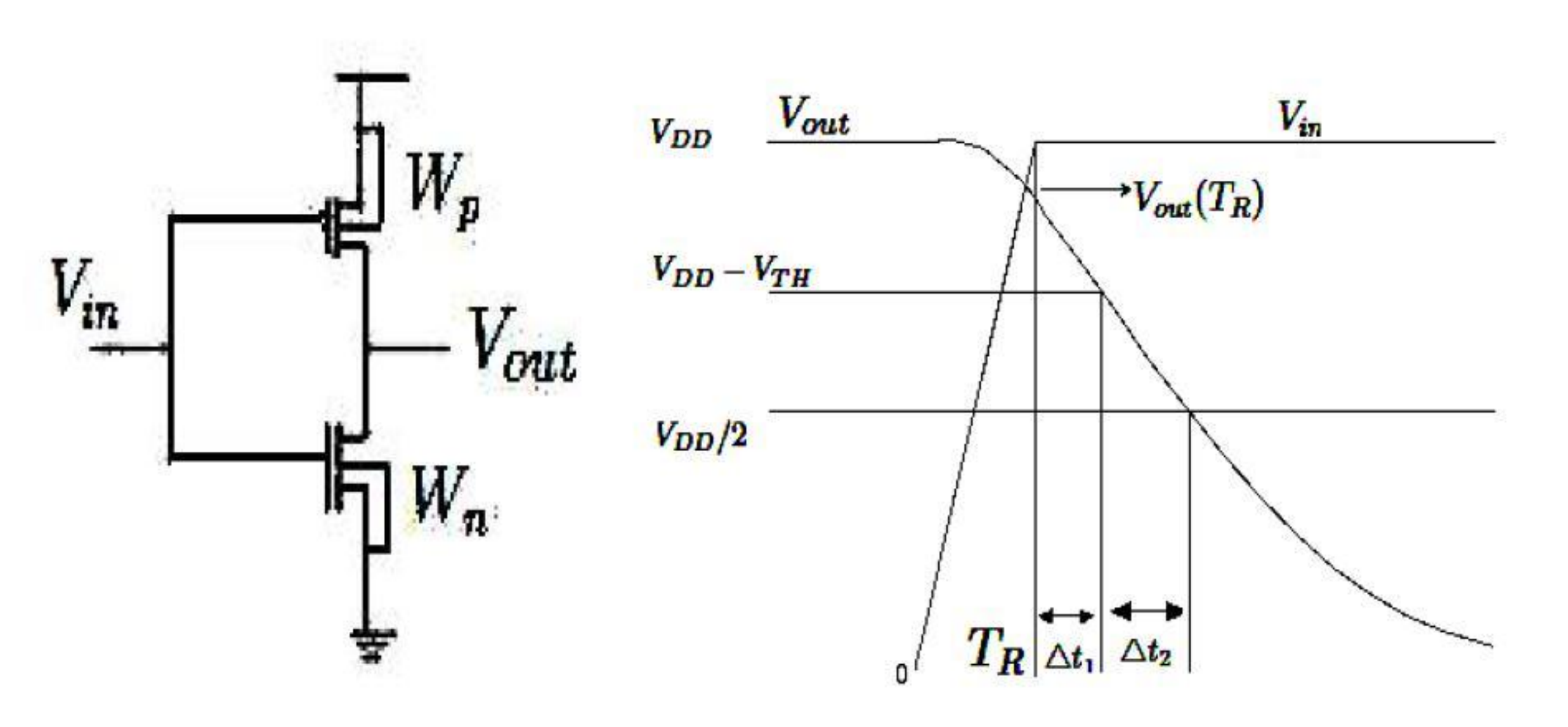}
	\caption{A CMOS Inverter and it's Input/Output}
	\label{cmos_not}
\end{figure}

For the NOT gate shown in the figure \ref{cmos_not} delay (between 50\% input to 50\% output) can be written as

\begin{equation}
\label{delay_def}
Delay= \frac{T_R}{2} + \Delta{}t_1+\Delta{}t_2
\end{equation}

Here we assume that when input reaches $V_{DD}$, $V_{out}$ drops a little and keeps NMOS in saturation. NMOS is also in saturation for time $T_R$ to $T_R + \Delta{}t_1$. NMOS falls into linear region from $T_R + \Delta{}t_1$ to $T_R + \Delta{}t_1 + \Delta{}t_2 $.
$V_{out}(T_R)$ can be calculated by charge discharged during time $0$ to $T_R$.

\begin{equation}
\Delta{}Q(T_R)=\int^{T_R}_0{}{I.dt}
\end{equation}

\begin{equation}
I= 0.5 \mu_nC_{ox}\frac{W}{L}(V_{GS}-V_{th})^2
\end{equation}

where
\begin{equation}
V_{GS}= \frac{V_{DD}}{T_R}t 
\end{equation}

\begin{equation}
\Delta{}Q(T_R) = \frac{1}{6}\mu_nC_{ox}\frac{W}{L}\frac{T_R}{V_{DD}}[(V_{DD}-V_{th})^3 +  (V_{th})^3 ]
\end{equation}

\begin{equation}
\Delta{}Q(T_R) = S_T T_R
\end{equation}

where 
\begin{equation}
S_T = \frac{1}{6}\mu_nC_{ox}\frac{W}{L}\frac{1}{V_{DD}}[(V_{DD}-V_{th})^3 +  (V_{th})^3 ]
\end{equation}

\begin{equation}
\Delta{}Q(T_R) = (C_L+C_P)(V_{DD}-V_{out}(T_R)))
\end{equation}

\begin{equation}
V_{out}(T_R) = V_{DD} -  \frac{\frac{1}{6}\mu_nC_{ox}\frac{W}{L}\frac{T_R}{V_{DD}}[(V_{DD}-V_{th})^3 +  (V_{th})^3 ]}{C_L+C_P}
\end{equation}

$\Delta{}t_1$ can be calculated by same method, but in this current will remain constant as $V_{GS}$ is constant and is equal to $V_{DD}$.

\begin{equation}
[V_{out}(T_R)- (V_{DD}- V_{th})] (C_L+C_P) = \int^{T_R+\Delta{}t_1}_{T_R}I_{const}.dt 
\end{equation}
where
\begin{equation}
I=0.5 \mu_nC_{ox}\frac{W}{L}(V_{DD}-V_{th})^2
\end{equation}

Since I is constant, integration simplifies as
\begin{eqnarray}
[V_{out}(T_R)- (V_{DD}- V_{th})] (C_L+C_P) \nonumber \\
 = 0.5 \mu_nC_{ox}\frac{W}{L}(V_{DD}-V_{th})^2 \Delta{}t_1
\end{eqnarray}

\begin{equation}
\label{delta_t1}
\Delta{}t_1 = \frac{(V_{th})(C_L+C_P)- S_T T_R}{0.5 \mu_nC_{ox}\frac{W}{L}(V_{DD}-V_{th})^2}
\end{equation}

Similarly we can calculate $\Delta{}t_2$, when NMOS is in linear region
\begin{equation}
\label{delta_t2}
\Delta{}t_2 = \alpha{}(C_L+C_P)\frac{1}{\mu_nC_{ox}\frac{W}{L}(V_{DD}-V_{th})} 
\end{equation}

Using Equation \ref{delay_def} , \ref{delta_t1}, \ref{delta_t2} we can write Delay as 

\begin{equation}
Delay = K_1 T_R + K_2 C_L + K_3
\end{equation}

where 

\begin{equation}
K_1= [0.5 - \frac{\frac{1}{6}\mu_nC_{ox}\frac{W}{L}\frac{1}{V_{DD}}[(V_{DD}-V_{th})^3 +  (V_{th})^3 ]}{0.5 \mu_nC_{ox}\frac{W}{L}(V_{DD}-V_{th})^2}]
\end{equation}

\begin{equation}
K_2= [ \frac{V_{th}}{0.5 \mu_nC_{ox}\frac{W}{L}(V_{DD}-V_{th})^2} + \frac{\alpha}{\mu_nC_{ox}\frac{W}{L}(V_{DD}-V_{th})} ]
\end{equation}

\begin{equation}
K_3= [ \frac{V_{th}C_P}{0.5 \mu_nC_{ox}\frac{W}{L}(V_{DD}-V_{th})^2} + \frac{\alpha{}C_P}{\mu_nC_{ox}\frac{W}{L}(V_{DD}-V_{th})} ]
\end{equation}

Our initial assumption was of NMOS being in saturation till time $T_R$. For this assumption to be valid, $V_{out}(T_R)$ should be greater than $V_{DD}-V_{th}$ i.e.  

\begin{equation}
V_{out}(T_R)  \geq  V_{DD}-V_{th}
\end{equation}

\begin{equation}
V_{DD} - \frac{S_T T_R}{C_L+C_P} \geq V_{DD} - V_{th}
\end{equation}

\begin{equation}
T_R \leq \frac{(C_L+C_P)V_{th}}{\frac{1}{6}\mu_nC_{ox}\frac{W}{L}\frac{1}{V_{DD}}[(V_{DD}-V_{th})^3 +  (V_{th})^3 ]}
\end{equation}

\begin{equation}
t_{rb} = \frac{(C_L+C_P)V_{th}}{\frac{1}{6}\mu_nC_{ox}\frac{W}{L}\frac{1}{V_{DD}}[(V_{DD}-V_{th})^3 +  (V_{th})^3 ]}
\end{equation}

\section{$t_{rb}$ Calculations for various Voltages and Technology corners}

\subsection{Variation with $C_L$}

In this section we will mathematically analyze the behavior of $t_{rb}$ with $C_L$.

\begin{equation}
t_{rb} = \frac{(C_L+C_P)V_{th}}{\frac{1}{6}\mu_nC_{ox}\frac{W}{L}\frac{1}{V_{DD}}[(V_{DD}-V_{th})^3 +  (V_{th})^3 ]}
\end{equation}

$t_{rb}$ is directly proportional to $C_L$, so we can write $t_{rb}$ as
\begin{equation}
\label{trb_linear_cl}
t_{rb}=M_1C_L+M_2
\end{equation}

where
\begin{equation}
\label{m_1}
M_1= \frac{V_{th}}{\frac{1}{6}\mu_nC_{ox}\frac{W}{L}\frac{1}{V_{DD}}[(V_{DD}-V_{th})^3 +  (V_{th})^3 ]}
\end{equation}

and
\begin{equation}
\label{m_2}
M_2= \frac{C_P V_{th}}{\frac{1}{6}\mu_nC_{ox}\frac{W}{L}\frac{1}{V_{DD}}[(V_{DD}-V_{th})^3 +  (V_{th})^3 ]}
\end{equation}

Now to model the exact behavior of $t_{rb}$, we need to know the behavior of constants $M_1$ and $M_2$ with supply voltage $V_{DD}$ and with chip  temperature.

\subsection{Variation with Supply Voltage}

On Simplifying Equation \ref{m_1} and \ref{m_2}, we get 

\begin{equation}
M_1= \frac{V_{th}}{\frac{1}{6}\mu_nC_{ox}\frac{W}{L}\frac{1}{V_{DD}}{V_{DD}^3}[(1-3 \frac {V_{th}}{V_{DD}}) +  (\frac{V_{th}}{V_{DD}})^3 ]}
\end{equation}

That can be further approximated. 
Since $(\frac{V_{th}}{V_{DD}})^3  \ll 1$ , $M_1$ can be written as
\begin{equation}
M_1\approx \frac{V_{th}}{\frac{1}{6}\mu_nC_{ox}\frac{W}{L}\frac{1}{V_{DD}}{V_{DD}^3}[(1-3 \frac {V_{th}}{V_{DD}}) ]}
\end{equation}

On further simplification
\begin{equation}
\label{m1_vdd}
M_1\approx \frac{V_{th}}{\frac{1}{6}\mu_nC_{ox}\frac{W}{L}V_{DD}^2}
\end{equation}

So $M_1$ is inversely proportional to $V_{DD}^2$.

Similarly $M_2$ can be characterized, as
\begin{equation}
\label{m2_vdd}
M_2\approx \frac{C_P V_{th}}{\frac{1}{6}\mu_nC_{ox}\frac{W}{L}V_{DD}^2}
\end{equation}

\subsection{Variation of $M_1$ and $M_2$ with chip temperature}

In the formula of $M_1$ in equation \ref{m_1}, parameters which are varying with temperature are mobility ($\mu_n$) and threshold voltage ($V_{th}$).

For Silicon, mobility of electrons  $\mu_n$  varies  with tempereture as follows [\ref{mobility_wiki}].

\begin{equation}
\label{mob_temp}
\mu_n \propto T^{-2.4}
\end{equation}

Similarly $V_{th}$ varies as  -3mv/$^o$C [\ref{vth_wiki}]
\begin{equation}
\label{vth_temp}
V_{th}(T_1) =V_{th}(T_2) - 0.003 \Delta T
\end{equation}

Where
\begin{equation}
\Delta T=  T_1 - T_2
\end{equation}

Combining equation [\ref{m1_vdd}, \ref{mob_temp}, \ref{vth_temp}], we see intuitively that while mobility increase $M_1$ and $M_2$ by a factor of $T^2.4$, $V_{th}$ reduces it to some extent. So we can roughly write it  as
\begin{eqnarray}
\label{m1_temp}
M_1 \propto T^2 \\
M_2 \propto T^2
\end{eqnarray}

\section{Observations through HPSICE simulations}

\begin{figure}[ht!]
	\centering
	\includegraphics[width=3.2in]{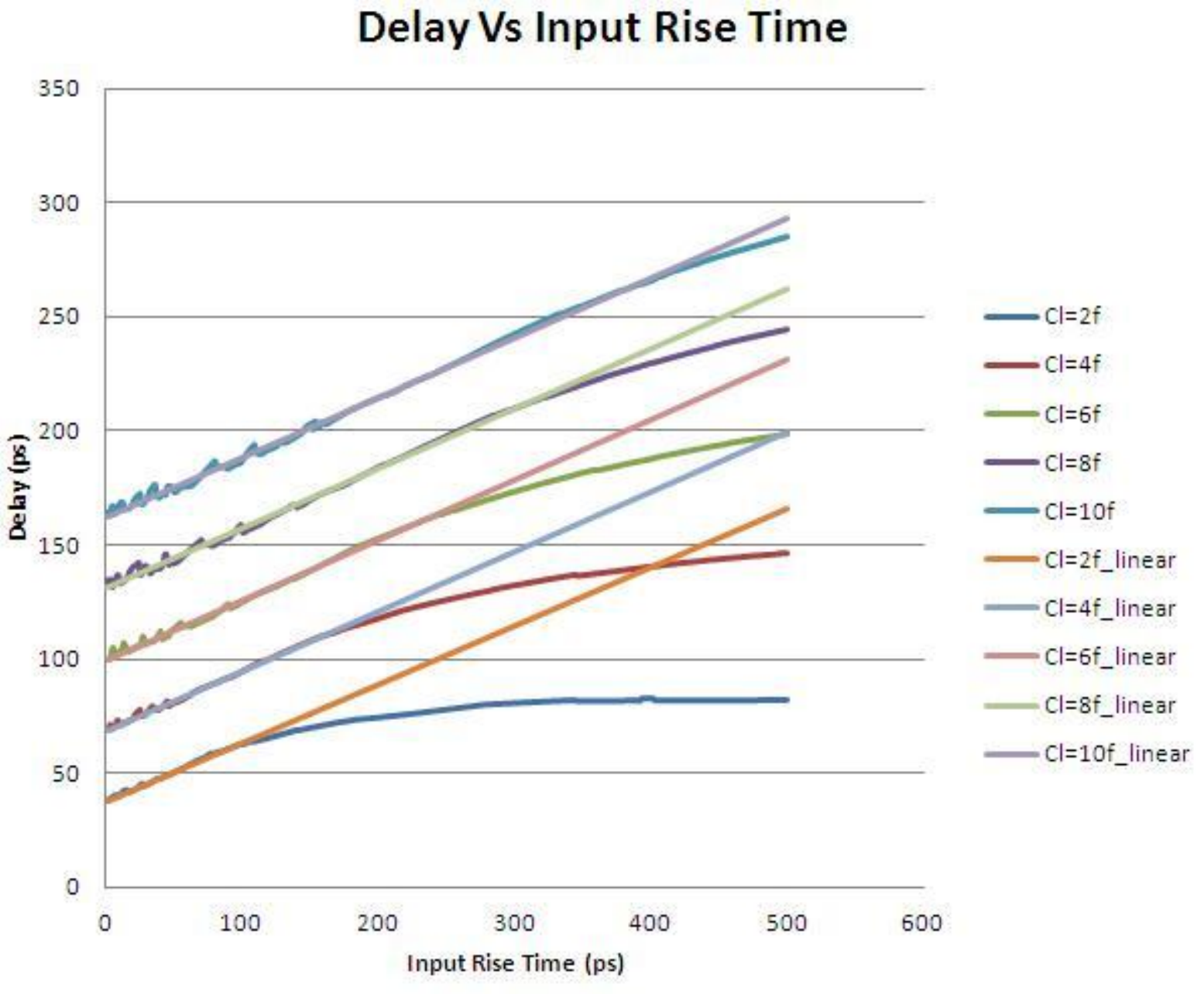}
	\caption{Delay vs $T_R$ at various $C_L$ and their linear regions}
	\label{delay_cl}
\end{figure}

We simulated CMOS NOT gate of figure \ref{cmos_not} using HSPICE at 45nm technology node. We collected values of delay (from 50\% input to 50\% output) by varying input rise time ($T_R$) from 1ps to 500ps at various load capacitances. Figure \ref{delay_cl} shows delay of NOT gate with respect to $T_R$ for various values of $C_L$.

Figure \ref{delay_cl} verifies that delay varies linearly up to certain extent. We captured this extent ($t_{rb}$) and plot it with various values of $C_L$.
\begin{figure}[ht!]
	\centering
	\includegraphics[width=3.2in]{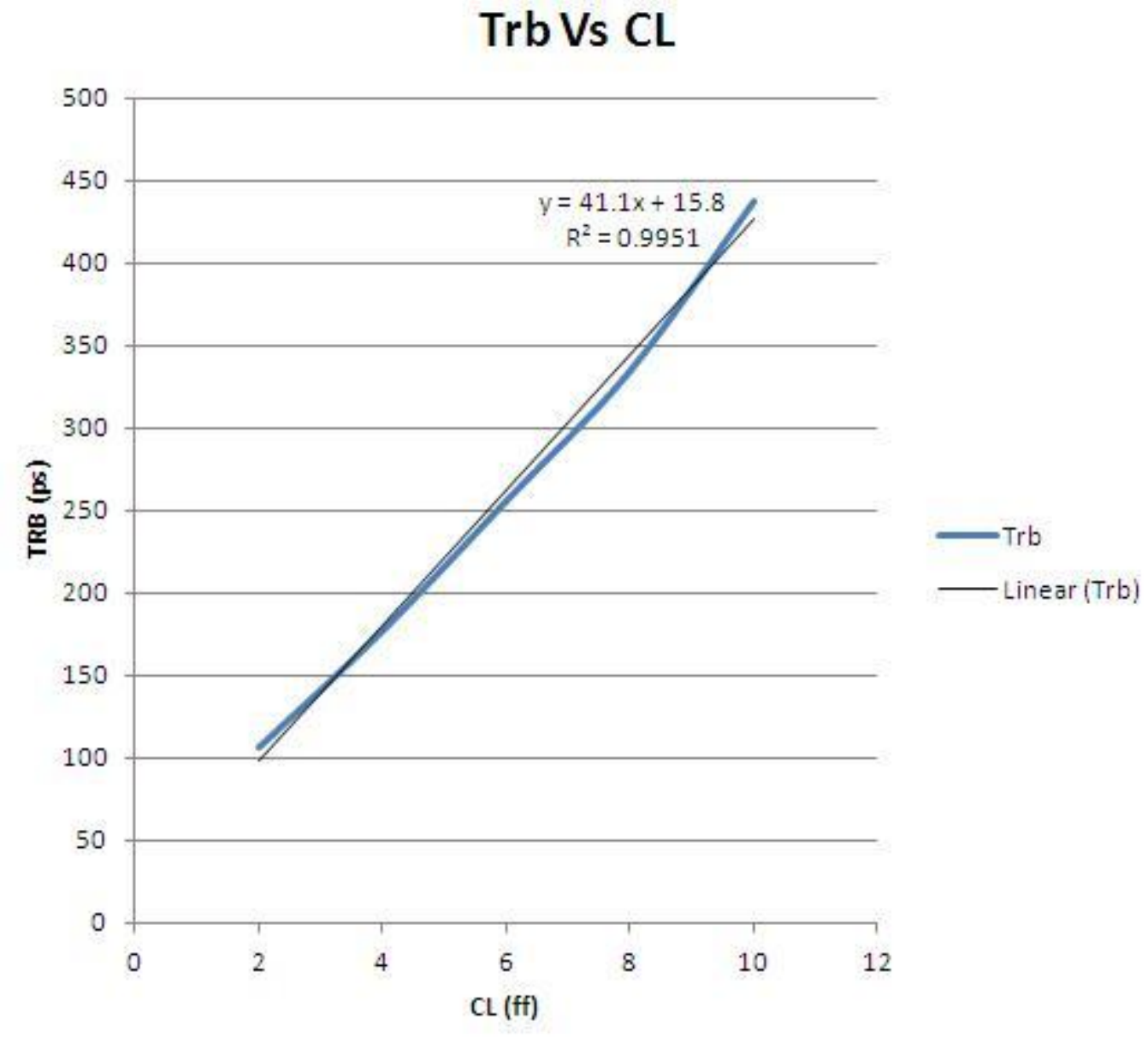}
	\caption{$t_{rb}$ vs $C_L$ and a Linear Fit on it}
	\label{trb_cl}
\end{figure}    

Figure \ref{trb_cl} verify our claims of equation \ref{trb_linear_cl} where we predicated $t_{rb}$ to vary linearly with $C_L$.We also plotted variations of $t_{rb}$ with supply voltage ($V_{DD}$).
\begin{figure}[ht!]
	\centering
	\includegraphics[width=3.2in]{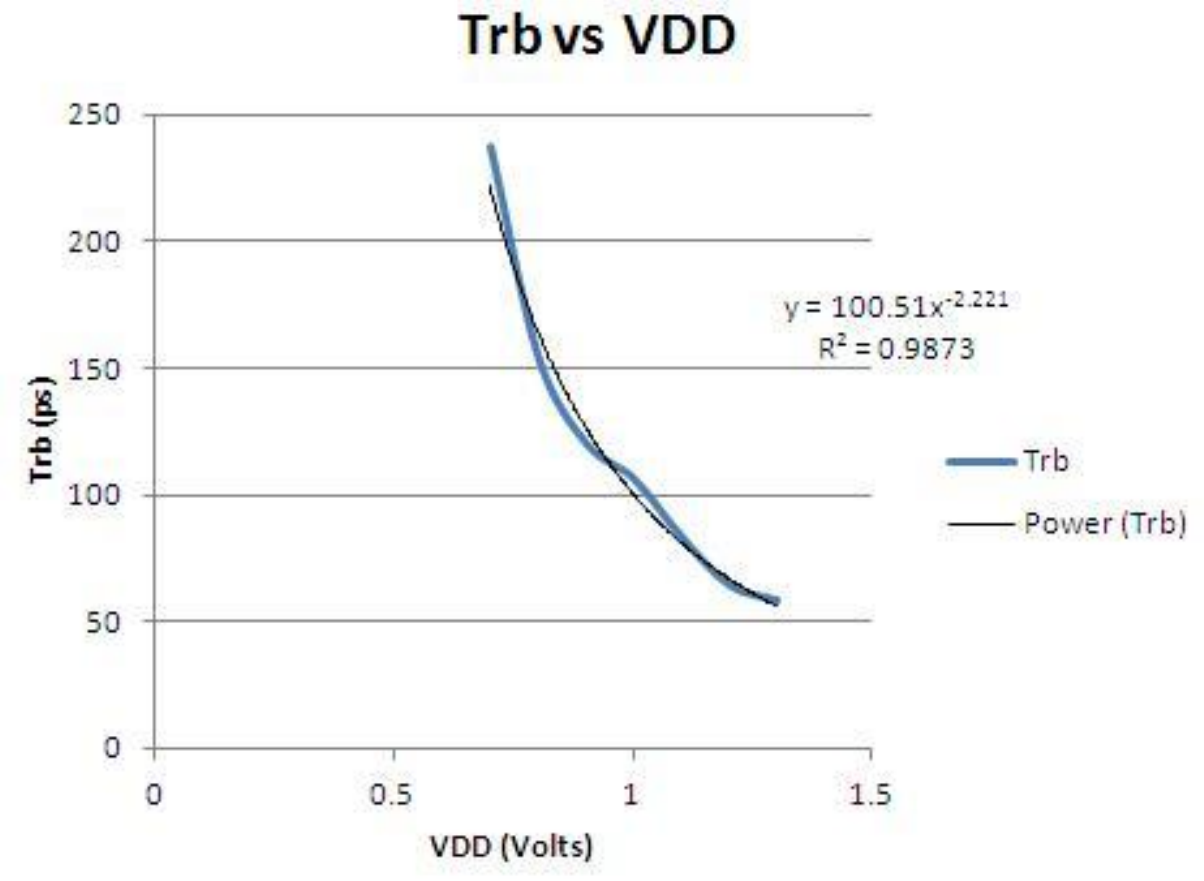}
	\caption{$t_{rb}$ vs $V_{DD}$ and a power function Fit on it}
	\label{trb_vdd}
\end{figure}

Since $M_1$ [Eq. \ref{m1_vdd}] and $M_2$ [Eq. \ref{m2_vdd}] both were inversely proportional to $V_{DD}^2$, $t_{rb}$ is also inversely proportional to $V_{DD}^2$. When we plotted simulation results of $t_{rb}$ vs $V_{DD}$ shown in figure \ref{trb_vdd},  we observed that it is varying in the predicated way. By fitting power function we observed that it varies as $V_{DD}^{-2.2}$, which is approximately equal to -2.

\begin{figure}[ht!]
	\centering
	\includegraphics[width=2.5in]{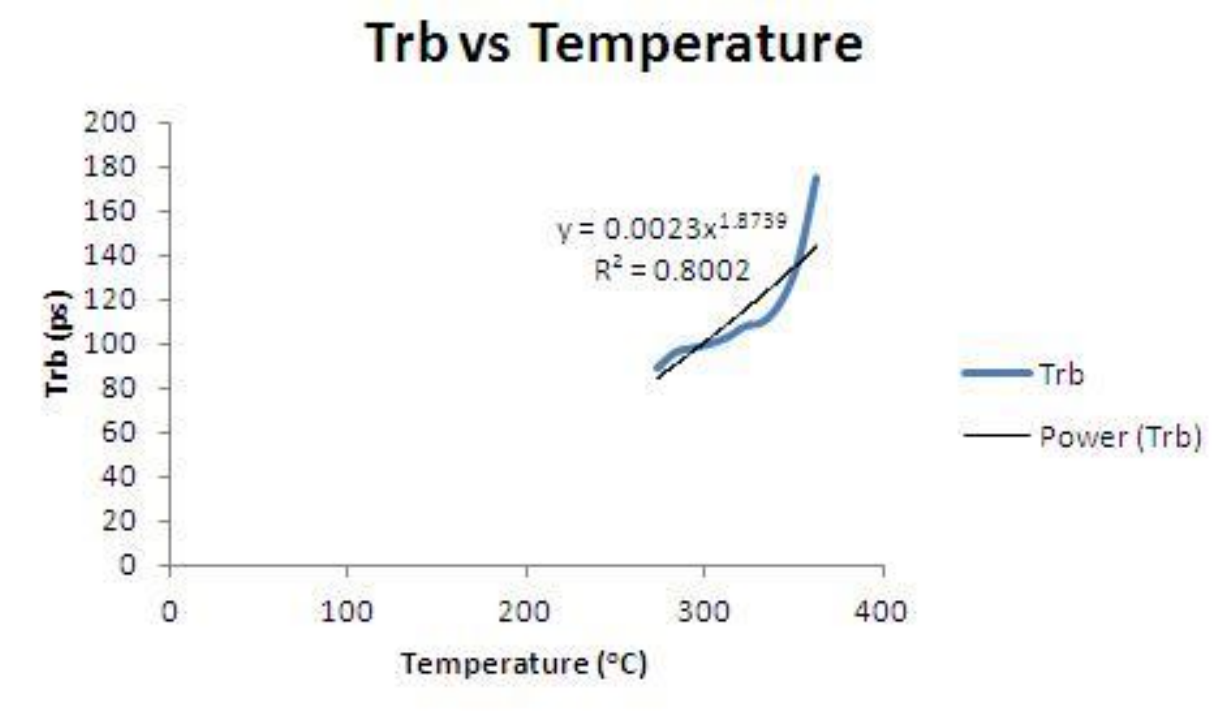}
	\caption{$t_{rb}$ vs $Temperature$ and a power function Fit on it}
	\label{trb_temp}
\end{figure}

In Equation \ref{m1_temp}, we proved that both $M_1$ and $M_2$ are proportional to $T^2$. Which implies same with $t_{rb}$. By Plotting $t_{rb}$ with temperature ($T$) shown in figure \ref{trb_temp}, we observed that $t_{rb}$ varies almost similar to the mathematical prove with $T^{1.87}$.

\section{Modelling $t_{rb}$}

For NOT gate we have seen the behavior of  $t_{rb}$ with Widths, Temperature, Supply Voltage and Load Capacitance.

\subsection{Adjusting $\frac{W}{L}$ variations}
from equation \ref{m_1}, \ref{m_2} it is clear that slope of $t_{rb}$ curve varies inversely with temperature while its intercept is constant. To model this, we will calculate two reference values of $t_{rb}$ at CL1 and CL2 and then we will model these for different widths.

\begin{equation}
Slope_{ref} = \frac{t_{rb\_ref2} - t_{rb\_ref1}}{C_{L2} - C_{L1}} 
\end{equation} 

\begin{equation}
Slope_{new} = \frac {Slope_{ref}*{(\frac{W}{L})_{ref}}}{(\frac{W}{L})_{new}} 
\end{equation} 

\begin{equation}
Intercept_{ref} =  \frac{C_{L1}*t_{rb\_ref2}-C_{L2}*t_{rb\_ref1}}{C_{L1} - C_{L2}} 
\end{equation} 

\begin{equation}
Intercept_{new} = \frac{C_{L1}*t_{rb\_ref2}-C_{L2}*t_{tb\_ref1}}{C_{L1} - C_{L2}} 
\end{equation}

So $t_{rb}$ at any width can be written as

\begin{eqnarray}
t_{rb} & = & \frac {Slope_{ref}*{(\frac{W}{L})_{ref}}}{(\frac{W}{L})_{new}} * C_L + \nonumber \\
& &  \frac{C_{L1}*t_{rb\_ref2}-C_{L2}*t_{rb\_ref1}}{C_{L1} - C_{L2}} 
\end{eqnarray}

\subsection{Adjusting $T$ variations}

From equation \ref{m1_temp}, it is clear that both slope and intercept varies in accordance with square of temperature. So $t_{rb}$ can be generalize as follow.

\begin{eqnarray}
t_{rb} & = & \frac {Slope_{ref}*{(\frac{W}{L})_{ref}}}{(\frac{W}{L})_{new}} * C_L + \nonumber \\
& & \frac{C_{L1}*t_{rb\_ref2}-C_{L2}*t_{rb\_ref1}}{C_{L1} - C_{L2}}* \frac{T_{new}}{T_{ref}} 
\end{eqnarray}

\subsection{Adjusting $V_{DD}$ variations}

From equation \ref{m1_vdd}, it is clear that both slope and intercept varies in accordance with inverse square of supply voltage. So $t_{rb}$ can be generalize as follow.

\begin{eqnarray}
t_{rb} & = & \frac {Slope_{ref}*{(\frac{W}{L})_{ref}}}{(\frac{W}{L})_{new}} * C_L + \nonumber \\
& & [\frac{C_{L1}*t_{rb\_ref2}-C_{L2}*t_{rb\_ref1}}{C_{L1} - C_{L2}}* \nonumber \\
& & \frac{T_{new}}{T_{ref}} * \frac{V_{DDref}}{V_{DDnew}}] 
\end{eqnarray}

\subsection{Adjusting Technology node variations}

From equation \ref{m1_vdd}, it is clear that both slope and intercept varies linearly with technology node. So $t_{rb}$ can be generalize as follow.

\begin{eqnarray}
t_{rb} & = & \frac {Slope_{ref}*{(\frac{W}{L})_{ref}}}{(\frac{W}{L})_{new}} * C_L + \nonumber \\
& & [\frac{C_{L1}*t_{rb\_ref2}-C_{L2}*t_{rb\_ref1}}{C_{L1} -C_{L2}}* \nonumber \\
& & (\frac{T_{new}}{T_{ref}})^2 * (\frac{V_{DDref}}{V_{DDnew}})^2 * \frac{L_{new}}{L_{ref}} ]
\end{eqnarray}

\section{LUT characterization using break point formula}
In previous sections we captured how $t_{rb}$ behaves with output load $C_L$, supply voltage $V_{DD}$ and on chip temperature variations. This formula tells that if we have $t_{rb}$ value at one corner, we can calculate it\rq{}s value at other node very easily. This approach will help us in determining linear region of delay curve at any $C_L$, $V_{DD}$ and temperature value. For LUT characterization, we will store the value of $t_{rb}$ for two  different corners and will calculate subsequent $t_{rb}$ using $t_{rb}$ model developed in previous sections. To get a feel of this, we can consider characterization of a look up table for NOT gate between $T_R$ range 1ps to 100ps and $C_L$ range 1ff to 10ff.

\begin{figure}[ht!]
	\centering
	\includegraphics[width=3in]{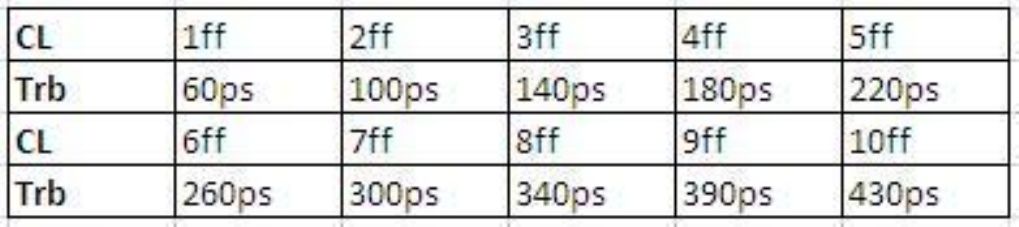}
	\centering \caption{$t_{rb}$ at various values of $C_L$}
	\label{trb_table}
\end{figure}

If we use traditional LUT characterization methods, we will divide this range into equal intervals. We will divide $T_R$ range into 20 points each of 5ps interval and $C_L$ range into 10 point each of 1ff interval, it will require 20x10 i.e . 200 SPICE simulations. With this new approach, we will calculate $t_{rb}$ values from our model at various $C_L$, which turns out as in figure \ref{trb_table}.

Figure \ref{trb_table} tells that we need only 2 simulations for $C_L$  having values 2ff to 10ff as whole region is linear, where as we will need 2+8 =10 simulations for $C_L$ of value 1ff as 60\% of region is linear. This in turn requires a total of only 28 (10+2x9=28) simulations. Thus we could save around 86\% of simulations.

\section{Conclusion}  
SPICE simulation at any particular corner has many advantage. Apart from giving delay it also gives power. In linear region we can do our LUT characterization without the need of simulations. But we will not have any idea about power consumption. So if power consumed is not our concern, demarcation of delay curve into two different region viz linear and non-linear helps us in fastening the process of LUT characterization. In future we can investigate similar kind of linear behavior in power consumption also which will obviate the need of SPICE simulation in linear region completely.

\end{document}